\documentclass[preprint,review,12pt]{elsarticle}

\usepackage{amssymb}
\usepackage[labelfont=bf, labelsep=space]{caption}
\usepackage[skip=0.5ex]{subcaption}
\usepackage{wrapfig}
\usepackage{xcolor, soul}
\usepackage{newtxtext,newtxmath,bm}
\newcommand\scalemath[2]{\scalebox{#1}{\mbox{\ensuremath{\displaystyle #2}}}}

\journal{Mathematical Biosciences}

\begin{document}

\begin{frontmatter}



\title{Antigenic Cooperation in Viral Populations: Redistribution of  Loads Among Altruistic Viruses and Maximal Load per Altruist}


\author[inst1]{Leonid Bunimovich}
\ead{leonid.bunimovich@math.gatech.edu}

\address[inst1]{School of Mathematics, Georgia Institute of Technology, Atlanta, 30332, Georgia, USA}

\author[inst1,inst2]{Athulya Ram \corref{cor1}}\ead{athulya@gatech.edu}

\address[inst2]{Interdisciplinary Graduate Program in Quantitative Biosciences, Georgia Institute of Technology, Atlanta, 30332, Georgia, USA}

\cortext[cor1]{Corresponding author}
\tnotetext[t1]{Declarations of interest: none}
\tnotetext[t2]{Acknowledgments: The authors acknowledge the support of Georgia Tech Interdisciplinary Graduate Program in Quantitative Biosciences. Funding: This work was partially supported by the NSF grant DMS-2054659.}

\begin{abstract}
The paper continues the study of the phenomenon of local immunodeficiency (LI) in viral cross-immunoreactivity networks, with a focus on the roles and interactions between altruistic and persistent viral variants. As always, only the state of stable (i.e. observable) LI is analysed. First, we show that a single altruistic viral variant has an upper limit for the number of persistent viral variants that it can support.  Our findings reveal that in viral cross-immunoreactivity networks, altruistic viruses act essentially autonomously from each other. Namely, connections between altruistic viruses do not change neither their qualitative roles, nor the quantitative values of the strengths of their connections in the CRNs. In other words, each altruistic virus does exactly the same actions and with the same strengths with or without presence of other altruistic viruses. However, having more altruistic viruses allows to keep sizes of populations of persistent viruses at the higher levels. Likewise, the strength of the immune response against any altruistic virus remains at the same constant level regardless of how many persistent viruses this altruistic virus supports, i.e. shields from the immune response of the host's immune system. It is also shown that viruses strongly compete with each other in order to become persistent in the state of stable LI. We also present an example for a CRN with stable LI that only consists of persistent viral variants.       
\end{abstract}

\begin{highlights}
\item \textbf{Autonomous Functioning of Altruistic Viruses:} This study demonstrates that altruistic viruses within cross-immunoreactivity networks (CRNs) operate independently of each other, maintaining stable local immunodeficiency regardless of their network connections. This autonomy highlights the robustness of altruistic viruses in supporting persistent viral variants.
\item \textbf{Maximum Load of Persistent Viruses:} We establish that the strength of the immune response against any altruistic virus remains at the same constant level regardless of how many persistent viruses this altruistic virus supports, i.e. shields from the immune response of the host's immune system. This shows that there is a maximum number of persistent variants that can be supported by a single altruistic viral variant.
\item \textbf{Cross-immunoreactivity networks with only persistent variants:} We present a general example of a type of CRN consisting only of persistent variants, while maintaining stable and robust local immunodeficiency (LI). This shows that altruistic variants, while helpful, are not necessary for the CRN to exhibit stable and robust LI.
\end{highlights}


\end{frontmatter}

%%\linenumbers

\section{Introduction}
The high mutation rates of RNA viruses such as Hepatitis C (HCV), HIV, Zika, Influenza A, and SARS-CoV-2 result in the existence of a heterogeneous population of genetically related viral variants within each infected host or community of infected individuals \cite{drake1999mutation, domingo2012viral}. This forms a complex ecosystem influenced by the dominant selection pressure caused by hosts' immune systems \cite{Rhee07}. Historically, the evolution of these viruses has been seen as an ``arms race," with the virus constantly accumulating genetic diversity to evade immune responses \cite{nowak_may2000}. However, recent experiments suggest a possibly more complex scenario. These studies reveal broad cross-immunoreactivity and antigenic convergence among viral variants within hosts \cite{campo2012hepatitis}, as well as increased negative selection and decreased population heterogeneity over time \cite{ramachandran2011temporal, campo2014next, gismondi2013dynamic, lu2008hcv, illingworth2014identifying}. Additionally, they show long-term persistence of viral variants \cite{ramachandran2011temporal, palmer2012insertion, palmer2014analysis} and fluctuating frequencies of subpopulations during infection \cite{ramachandran2011temporal, gismondi2013dynamic, palmer2014analysis, gray2012new, raghwani2016exceptional}. These findings suggest that viral evolution is driven by multiple complex mechanisms rather than a single process. It is a non-linear dynamics influenced by successive selection challenges specific to different stages of infection or epidemic spread \cite{baykal2020quantitative}, involving mechanisms common across various life forms \cite{domingo2019social, baykal2020quantitative}.

Several previous mathematical models of virus-host immune interactions typically suggest that immune escape is characterized by a continuous increase in genomic diversity \cite{nowak_may2000, wodarz2003hepatitis, iwasa2004some}. However, these models do not account for several experimentally observed phenomena, including the transition from immune escape under positive selection in the early stages of the infection to a conditionally stable state under negative selection later on. Some studies have linked this phenomenon to cross-immunoreactivity, explaining prolonged stasis in immune escape and the coexistence of intra-host variant clusters \cite{haraguchi1997evolutionary, gog2002dynamics}. More recent studies propose that specific cooperative interactions among viral variants \cite{pnas, shirogane2013cooperation, domingo2019social, baykal2020quantitative} allow viral populations to adapt as quasi-social systems \cite{domingo2019social}.

Our previous studies \cite{pnas, BS1, BS2, bunimovich2019specialization, BuRam1, BuRam2} introduced and analyzed a nonlinear  ODE model describing interactions between viral antigens and host B cells, with a central role played by the cross-immunoreactivity network (CRN) of viral antigens. Unlike traditional models, this model considers antigenicity (the ability to bind antibodies) and immunogenicity (the ability to elicit antibodies) as separate factors \cite{campo2012hepatitis, van2012basic, freitas1995role, mclean1997resource, tarlinton2006b, schwickert2007vivo, Pal12}. The complex topology of the CRN creates a dynamic fitness landscape where viral variants influence each other's fitness. This setup leads to antigenic cooperation and specialization, where certain variants (altruists) sacrifice their fitness to boost the fitness of others (selfish/persistent variants), resembling kin selection \cite{hamilton1964genetical}. This model explains various empirical observations and maintains a stable and robust state of LI under realistic conditions \cite{BS1}. For LI, stability suggests that small changes in initial conditions lead to the same stable state, while robustness suggests that small changes in parameters do not change the existence of stable state of LI. Notably, the model in \cite{pnas} achieves predictive power with fewer variables compared to other models \citep{wodarz2003,nowak_etal1990, nowak_etal1991, nowak_may1991} due to its high non-linearity and detailed exploration of cross-immunoreactivity effects.

The initial study \cite{pnas} introduced the antigenic cooperation model, highlighting its ability to explain local immunodeficiency and antigenic cooperation through numerical simulations and  analysis of equilibrium states. It was demonstrated in \citep{BS1} that local immunodeficiency solutions are stable and robust across various conditions for specific CRN types. Further, in the study \citep{BS2}, we explored the role of altruistic variants in intra-host adaptation, showing that without them, the viral population could only achieve marginal stability and smaller size.

Viral populations and CRNs are dynamic, subject to changes due to the emergence or introduction of variants with new phenotypes. This raises critical questions about how changes in CRNs drive evolutionary transitions and affect immune escape. In the study \cite{BuRam1}, the focus was on the impact of new variant emergence and viral transmission between chronically infected hosts on B cell immune adaptation. These processes were analyzed by first examining the new stable state post-emergence or transmission and then studying the dynamic transition from the initial to the new state. The study showed that these events can rapidly rearrange the viral ecosystem and alter the roles of viral variants. The study also demonstrated that new antigenic variants can coexist with persistent ones, maintaining local immunodeficiency in the CRN. Unexpectedly, new variants can also shift the population from immune escape through genomic diversification to stable adaptation via antigenic cooperation. These insights emphasized the phenotypic influence of both antibody and quasi-social environments on viral variants, rather than just their genomes. This complexity poses challenges for effective vaccine design, as the introduction of new variants significantly impacts the evolutionary trajectories of viral populations.

The current study builds on these foundational models by introducing a new example of a cross-immunoreactivity network that demonstrates stable and robust LI. This study aims to explore how the structure and dynamics of CRNs influence viral persistence, focusing on the roles of altruistic and persistent viral variants. In all previous studies, altruistic variants were believed to be an integral part of a CRN, without them stable and robust LI could not be formed. However, in this study, we introduce the first CRN which exhibits stable and robust LI that consists of only persistent variants.

Persistent variants gain fitness at the expense of altruistic variants. On average, the number of persistent viruses exceeded the number of altruistic viruses by about ten times in about eight hundred computer simulations detailed in \cite{pnas}. This naturally leads us to the question whether there is a maximal load of persistent variants that can be supported by an altruistic variant, and whether altruistic variants cooperate with each other to support more persistent variants. In this study, we find the upper limit of persistent variants that can be supported by a single altruistic variant in a very simple CRN. Then we explore small CRNs with various symmetries to study whether two altruists cooperate with each other.

By analyzing these interactions, we seek to deepen our understanding of the evolutionary strategies employed by RNA viruses and their implications for immune adaptation and vaccine design.

\section{A Mathematical Model of Evolution of Viral Diseases with Non-homogeneous cross-immunoreactivity Networks}\label{sec: model}

In this section, we will describe the mathematical model of evolution of diseases with non-homogeneous CRNs. Even though this model was introduced in \cite{pnas} for Hepatitis-C, this can be applied for any disease that exhibits cross-immunoreactivity.

Cross-immunoreactivity of viral variants is described by a directed and weighted graph $G_{CRN} = (V, E)$ with viral variants at vertices and edges that connect variants with cross-immunoreactivity. The population of viral variants is denoted by $x_i$ and they induce immune responses in the form of antibodies (Abs), denoted by $r_i$. 

Not every interaction of a virus with an antibody leads to neutralization. To factor this interaction, we assign weight functions to the network given by two matrices $U = (u_{ij})_{i,j = 1}^n$ and $V = (v_{ij})_{i,j = 1}^n$. The matrix $U$ corresponds to the ability of an immune response to neutralize a viral variant, and is given by $U = \text{Id} + \beta A^T$ where $\beta$ is the immune stimulation constant and $A$ is the adjacency matrix of $G_{CRN}$. Similarly, the matrix $V$ corresponds to the ability of a viral variant to stimulate an immune response, given by $V = \text{Id} + \alpha A$ where $\alpha$ is the neutralization coefficient. It is assumed that $u_{ii} = v_{ii} = 1$, i.e. the immune response against variant $i$ always neutralizes variant $i$, and viral variant $i$ always completely stimulate immune response $i$. The neutralization of a virus (antigen) may need more than one antibody, so we consider $0 < \beta = \alpha^k < \alpha < 1$.

The evolution of the viral variants and their corresponding immune responses is described using the following system of ordinary differential equations \cite{pnas}:

\begin{equation}\label{population}
\begin{split}
\dot x_i=f_ix_i-px_i\sum_{j=1}^nu_{ji}r_j,\quad i=1,\dots,n,\\
\dot r_i=c\sum_{j=1}^nx_j\frac{v_{ji}r_i}{\sum_{k=1}^nv_{jk}r_k}-br_i,\quad i=1,\dots,n.
\end{split}
\end{equation}

The viral variant of population $x_i$ has a replication rate of $f_i$ and it is eliminated by the immune responses $r_i$ at the rate $p u_{ji} r_j$, where $p$ is a constant. As for the growth of the immune response, the viral variant $x_j$ stimulates the immune response $r_j$ at the rate $c g_{ji} x_j$, where $g_{ji} = \frac{v_{ji}r_i}{\sum_{k=1}^nv_{jk}r_k}$ is the probability of the corresponding stimulation, and $c$ is a constant. The immune response is cleared at a constant rate $b$.

This model \cite{pnas} allows us to integrate the phenomenon of the original antigenic sin, which states that the viral variant $x_i$ preferentially stimulates preexisting immune responses that are capable of binding to $x_i$. In this paper we consider a more general model than the one studied in \cite{pnas} and assume that the viruses can have different replication rates.

In all papers dealing so far with a local immunodeficiency \cite{pnas, BS1, BS2, BuRam2} a viral variant was defined as persistent if the concentration of its immune response became lower than its initial value (in numerical simulations),
or if this immune response was completely eliminated (in mathematical studies), while maintaining a non-zero population. Similarly, a viral variant was defined as altruistic if the concentration of its population became lower than its initial value (in numerical simulations), or if the population was completely eliminated (in mathematical studies), while maintaining a non-zero immune response against it.\\

While introducing the mathematical model, \cite{pnas} only focused on persistent and altruistic viral variants and simply categorized the rest of the variants as ``others". In \cite{BS1, BS2, BuRam2} these other variants were named ``neutral active" and ``neutral inactive". Neutral active variants existed in a non-zero population with non-zero concentration of immune responses against them, and neutral inactive variants were practically non-existent in the system with a zero population and zero immune response against them.\\

In \cite{BuRam1}, persistent viruses were defined to have an equilibrium immune response strength $r_i^* \leq r_i^{\circ}$, where $r_i^{\circ} = \frac{f_i}{p}$, which is the strength of the immune response in the absence of cross-immunoreactivity. This new definition eliminated the category ``neutral active", and the viral variants which were initially called ``neutral inactive" were renamed as ``transient". Transient variants has populations and immune responses that asymptotically reaches zero, which explains the name as these variants are only present in the host system temporarily.\\

In this section, we are revisiting the thresholds that categorize viral variants into the categories - persistent, altruistic, neutral and transient.\\

In the mathematical model \ref{population}, if we assume that there is no cross-immunoreactivity, then the model reduces to:

\begin{equation}\label{reduced}
\begin{split}
\dot x_i=f_ix_i-px_ir_i,\quad i=1,\dots,n,\\
\dot r_i=c x_i-br_i,\quad i=1,\dots,n.
\end{split}
\end{equation}

The system reduces to the model in \cite{nowak_may2000}. From \ref{reduced}, the immune response against a virus in the absence of CR is $r_i^{\circ} = \frac{f_i}{p}$.\\

This means that if there was no CR in the model, all viral variants will have an immune response of $r_i^{\circ} = \frac{f_i}{p}$ against them. As persistent variants take advantage of CR to reduce the immune response against them (effectively hiding from the host immune system), we can categorize variants with a non-zero population and immune response strength $r_i^* < r_i^{\circ}$ as persistents. Similarly, as altruistic viral variants draw a majority of the host's immune response towards them, these variants would have an immune response strength $r_i^* > r_i^{\circ}$.\\

Therefore, we can define the state of local immunodeficiency as an equilibrium solution \textbf{$(x^*, r^*)$} such that every viral variant $i$ falls into one of the four following categories:
\begin{itemize}
\item[1)] $x_i^* > 0$ and $r_i^* < r_i^{\circ}$ ({\it persistent variants});

\item[2)] $x_i^* = 0$ and $r_i^* > r_i^{\circ}$ ({\it altruistic variants});

\item[3)] $x_i^* = r_i^* = 0$ ({\it transient variants});

\item[4)] $x_i^* > 0$ and $r_i^* = r_i^{\circ}$ ({\it neutral variants}).
\end{itemize}

\section{A new example of a cross immunoreactivity network with stable and robust local immunodeficiency}
\begin{figure}[ht]
    \centering
    \includegraphics[scale=0.3]{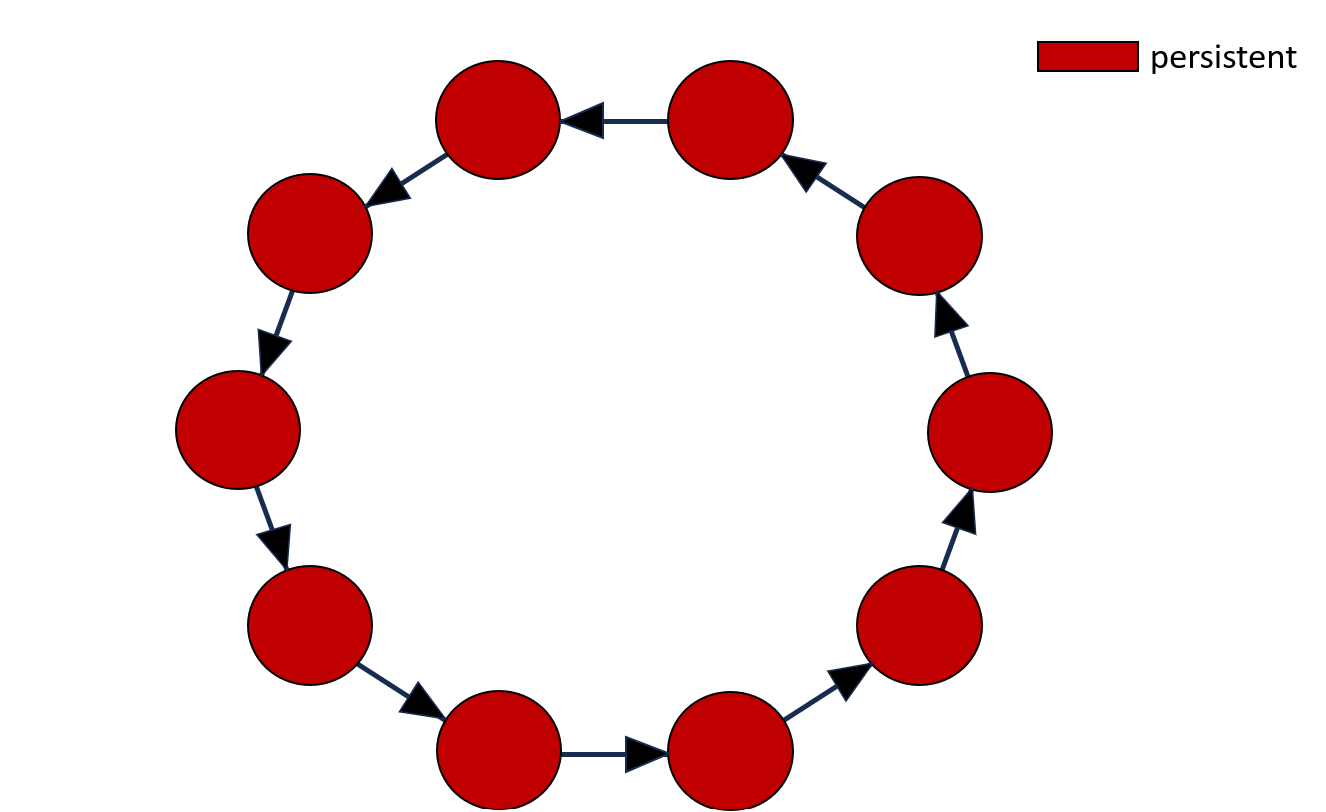}
    \caption{A simple cyclic network that exhibits stable local immunodeficiency}
    \label{fig:simple_cyclic_10}
\end{figure}

From the definition in section \ref{sec: model}, we present a new example for a network that exhibits stable and robust local immunodeficiency, shown in figure \ref{fig:simple_cyclic_10}. This is a simple cyclic network of $n$ viral variants. In this network, we assume that the viral variants have the same replication rate $f$.

At equilibrium, the strength of the immune responses of this network is $r_i^* = \frac{f}{p \sum_{j=1}^n U_{ji}}$ from equation \ref{population}. As $\sum_{j=1}^n U_{ji} > 1$ ($U =$ Id $+ \beta A^T$ and hence $u_{ii} = 1$), we have $\frac{f}{p \sum_{j=1}^n U_{ji}} < \frac{f}{p}$, which gives $r_i^* < r_i^{\circ}$, and thus makes all the $n$ viral variants in the network persistent.\\

\begin{figure}[ht]
    \centering
    \includegraphics[scale=0.3]{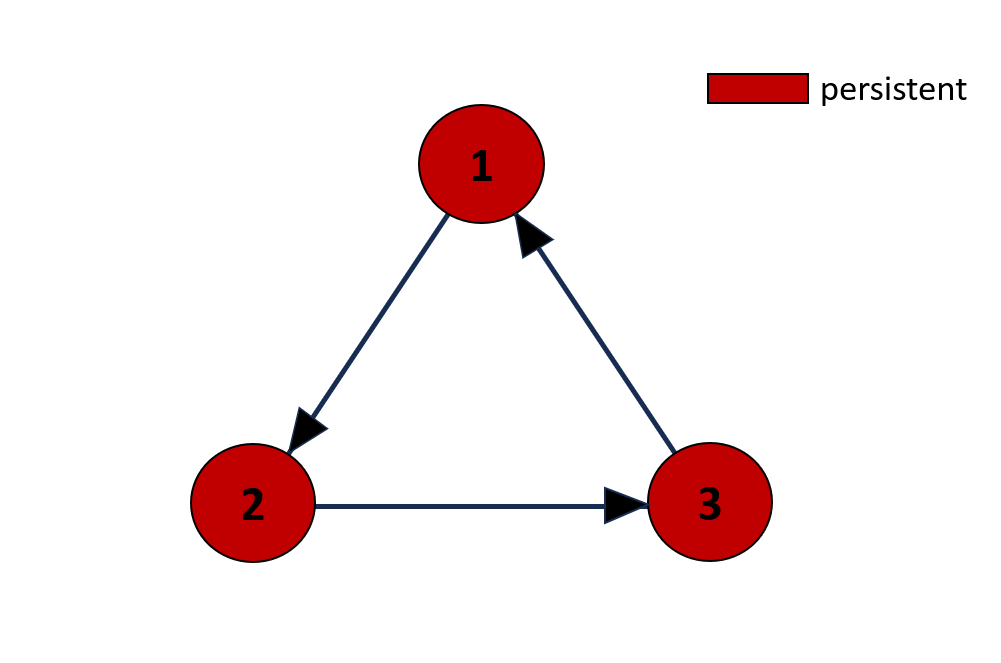}
    \caption{Simple cyclic network with 3 viral variants}
    \label{fig:3-cycle}
\end{figure}

Looking at the most simple version of this network with just three viral variants, we can prove the stability of the equilibrium. Consider the network shown in figure \ref{fig:3-cycle}. For this network, corresponding matrices $U$ and $V$ are:

\begin{align*}
    U = \begin{pmatrix}
    1 &0 &\beta\\
    \beta &1 &0\\
    0 &\beta &1
\end{pmatrix}, & \quad &
V = \begin{pmatrix}
    1 &\alpha &0\\
    0 &1 &\alpha\\
    \alpha &0 &1
\end{pmatrix}.
\end{align*}
The governing equations for this network are:

\begin{flalign}\label{eqn:3-cycle}
\begin{split}
    \dot{x_1} &= f x_1 - p x_1 (r_1 + \beta r_2 ),\\
    \dot{x_2} &= f x_2 - p x_2 (r_2 + \beta r_3),\\
    \dot{x_3} &= f x_3 - p x_3 (r_3 + \beta r_1),\\
    \dot{r_1} &= c(\frac{x_1 r_1}{r_1 + \alpha r_2} + \frac{\alpha x_3 r_1}{r_3 + \alpha r_1}) - b r_1,\\
    \dot{r_2} &= c(\frac{\alpha x_1 r_2}{r_1 + \alpha r_2} + \frac{x_2 r_2}{r_2 + \alpha r_3}) - b r_2,\\
    \dot{r_3} &= c(\frac{\alpha x_2 r_3}{r_2 + \alpha r_3} + \frac{x_3 r_3}{\alpha r_1 + r_3}) - b r_3.
\end{split}
\end{flalign}

The fixed point that exhibit local immunodeficiency is:
\begin{flalign*}
    x_1^* &= \frac{b(f + (\alpha - \beta)(f - \beta f) + \alpha \beta^2 f)}{cp(\beta^3 +1)(\alpha + 1)}, &r_1^* &= \frac{f}{p(\beta + 1)},\\
    x_2^* &= \frac{b(f + (\alpha - \beta)(f - \beta f) + \alpha \beta^2 f)}{cp(\beta^3 +1)(\alpha + 1)}, &r_2^* &= \frac{f}{p(\beta + 1)},\\
    x_3^* &= \frac{b(f + (\alpha - \beta)(f - \beta f) + \alpha \beta^2 f)}{cp(\beta^3 +1)(\alpha + 1)}, &r_3^* &= \frac{f}{p(\beta + 1)}.
\end{flalign*}

In the 3-cycle network, $\sum_{j=1}^n U_{ji} = \beta + 1$ for $i = 1,2,3$. Therefore $r_i^* = \frac{f}{p \sum_{j=1}^n U_{ji}} < r_i^{\circ}$, hence all three viral variants are persistent. The stability of this fixed point is explained in \ref{3-cycle}.

At the steady state of equation \ref{eqn:3-cycle}, we have 

\begin{flalign*}
    f = p(r_1^* + \beta r_2^*)\\
    f = p(r_2^* + \beta r_3^*)\\
    f = p(r_3^* + \beta r_1^*)\\
    b = c(\cfrac{x_1^*}{r_1^* + \alpha r_2^*} + \cfrac{\alpha x_3^*}{\alpha r_1^* + r_3^*})\\
    b = c(\cfrac{\alpha x_1^*}{r_1^* + \alpha r_2^*} + \cfrac{x_2^*}{\alpha r_3^* + r_2^*})\\
    b = c(\cfrac{\alpha x_2^*}{r_2^* + \alpha r_3^*} + \cfrac{x_3^*}{\alpha r_1^* + r_3^*})
\end{flalign*}

From $r_1^* < \frac{f}{p}$,

\begin{flalign*}
    &r_1^* + \alpha r_2^* < \frac{f}{p} + \alpha \frac{f}{p} = \frac{f + \alpha f}{p}\\
    &\implies \frac{1}{r_1^* + \alpha r_2^*} > \frac{p}{f(1 + \alpha)}  
\end{flalign*}

Similarly, $\cfrac{1}{r_3^* + \alpha r_1^*} > \cfrac{p}{f(1 + \alpha)}$ and $\cfrac{1}{r_2^* + \alpha r_3^*} > \cfrac{p}{f(1 + \alpha)}$.

\begin{flalign*}
    \frac{b}{c} &= \cfrac{x_1^*}{r_1^* + \alpha r_2^*} + \cfrac{\alpha x_3^*}{\alpha r_1^* + r_3^*}\\
    &> \cfrac{p x_1^*}{f(1 + \alpha)} + \cfrac{\alpha p x_3^*}{f(1 + \alpha)}\\
    &= \cfrac{p(x_1^* + \alpha x_3^*)}{f(1 + \alpha)}\\
    &\implies \cfrac{bf(1 + \alpha)}{cp} > x_1^* + \alpha x_3^*
\end{flalign*}
Similarly, $\cfrac{bf(1 + \alpha)}{cp} > x_2^* + \alpha x_1^*$ and $\cfrac{bf(1 + \alpha)}{cp} > x_3^* + \alpha x_2^*$.\\

In the cyclic network, as $f = f_1 = f_2 = f_3$, we also have $x^* = x_1^* = x_2^* = x_3^*$.

\begin{flalign*}
    &x^* + \alpha x^* < \cfrac{bf(1 + \alpha)}{cp}\\
    &\implies (1 + \alpha) x^* < \cfrac{bf(1 + \alpha)}{cp}\\
    &\implies x^* < \cfrac{bf}{cp}.
\end{flalign*}

This can be extended to $n$-cycle networks. Therefore for circular directional networks, $r^* < \cfrac{f}{p}, x^* < \cfrac{bf}{cp}$ when all variants have the same replication rates $f$.

\section{Maximal load on a single altruistic virus}
\begin{figure}[ht]
    \centering
    \includegraphics[width=0.5\textwidth]{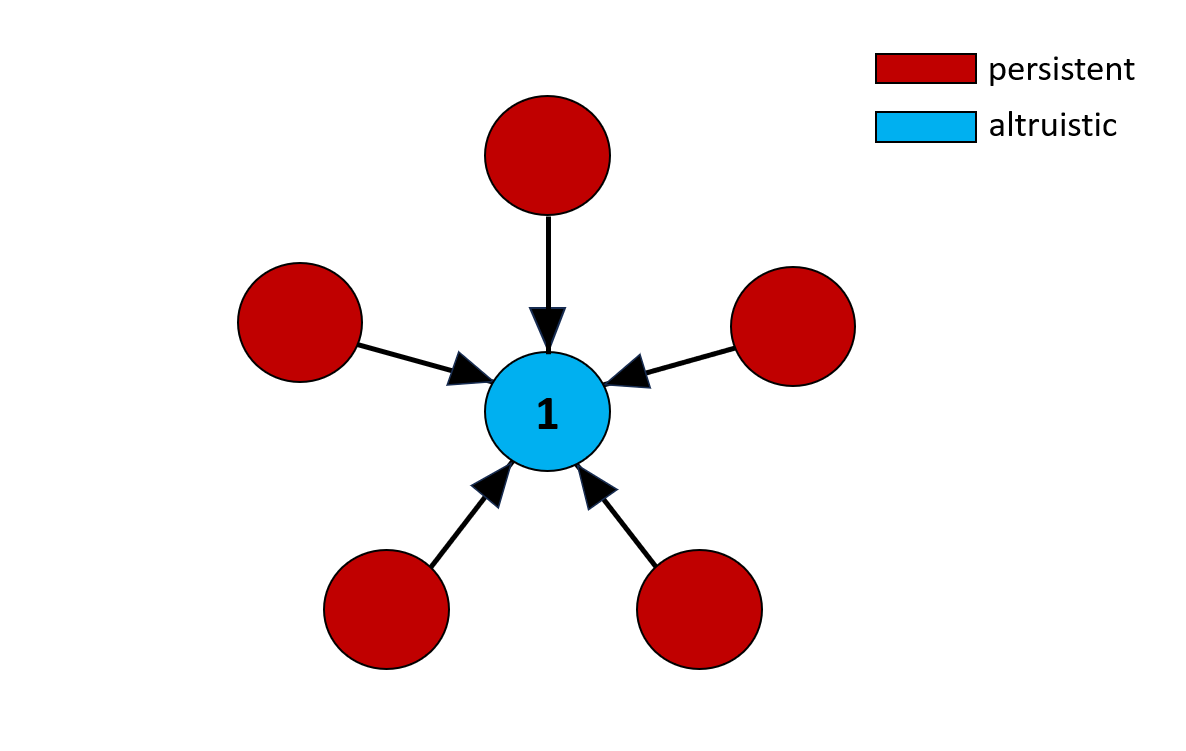}
    \caption{A basic network structure with one altruistic virus and multiple persistent viruses}
    \label{fig:simple_structure_one_altruist}
\end{figure}

In this section, we will determine a maximum number of persistent viruses that can be supported by a single altruistic virus. This is a basic problem for understanding the structure of a cross-immunoreactivity network which demonstrates the phenomenon of local immunodeficiency. So far, all existing computer simulations, starting with \cite{pnas}, demonstrated that (in large CR networks) many persistent viruses were connected to each altruistic virus. For instance, in about eight hundred simulations the number of persistent viruses exceeded (on average) the number of altruistic viruses by about ten times. Moreover, persistent viruses always were very ``selfish" in the sense that they did not have any connections between themselves trying to extract a maximal support from altruistic viruses without sharing it with other persistent viruses, but rather competing with each other. Therefore in order to make the first step to understanding what is a maximal load which an altruistic virus can sustain we consider a simple network where a finite number of persistent viruses are connected to one and the same altruist (figure \ref{fig:simple_structure_one_altruist}).

It is known \cite{BuRam1, BuRam2} that when viruses having different replication rates are connected to an altruistic virus, then the viruses compete with each other. As a result only the viruses with a maximal replication rate become persistent, while all other viruses connected to an altruistic virus disappear (become transient). Therefore all persistent viruses in a case such as \ref{fig:simple_structure_one_altruist}, supported by the same altruistic virus, should have one and the same replication rate ($f$).

We start with analysis of the simplest situation when in the network depicted in figure \ref{fig:simple_structure_one_altruist} there are only two persistent viruses. It is assumed that both of them have the same replication rate, otherwise the ``weaker" one can not remain to be persistent.

Then the governing equations of the dynamics (evolution) of this system have the following form.

\begin{flalign}\label{eqn:3network_ode}
\begin{split}
    \dot{x_1} &= f_a x_1 - p x_1 r_1,\\
    \dot{x_2} &= f_p x_2 - p x_2 (\beta r_1 + r_2),\\
    \dot{x_3} &= f_p x_3 - p x_3 (\beta r_1 + r_3),\\
    \dot{r_1} &= c(x_1 + \frac{\alpha x_2 r_1}{\alpha r_1 + r_2} + \frac{\alpha x_3 r_1}{\alpha r_1 + r_3}) - b r_1,\\
    \dot{r_2} &= c(\frac{x_2 r_2}{\alpha r_1 + r_2}) - b r_2,\\
    \dot{r_3} &= c(\frac{x_3 r_3}{\alpha r_1 + r_3}) - b r_3.
\end{split}
\end{flalign}
where $f_a$ is the replication rate of the altruistic virus and $f_p = f_2 = f_3$ is the replication rate of the persistent viruses.\\

\begin{figure}[ht]
    \centering
    \includegraphics[width=\textwidth]{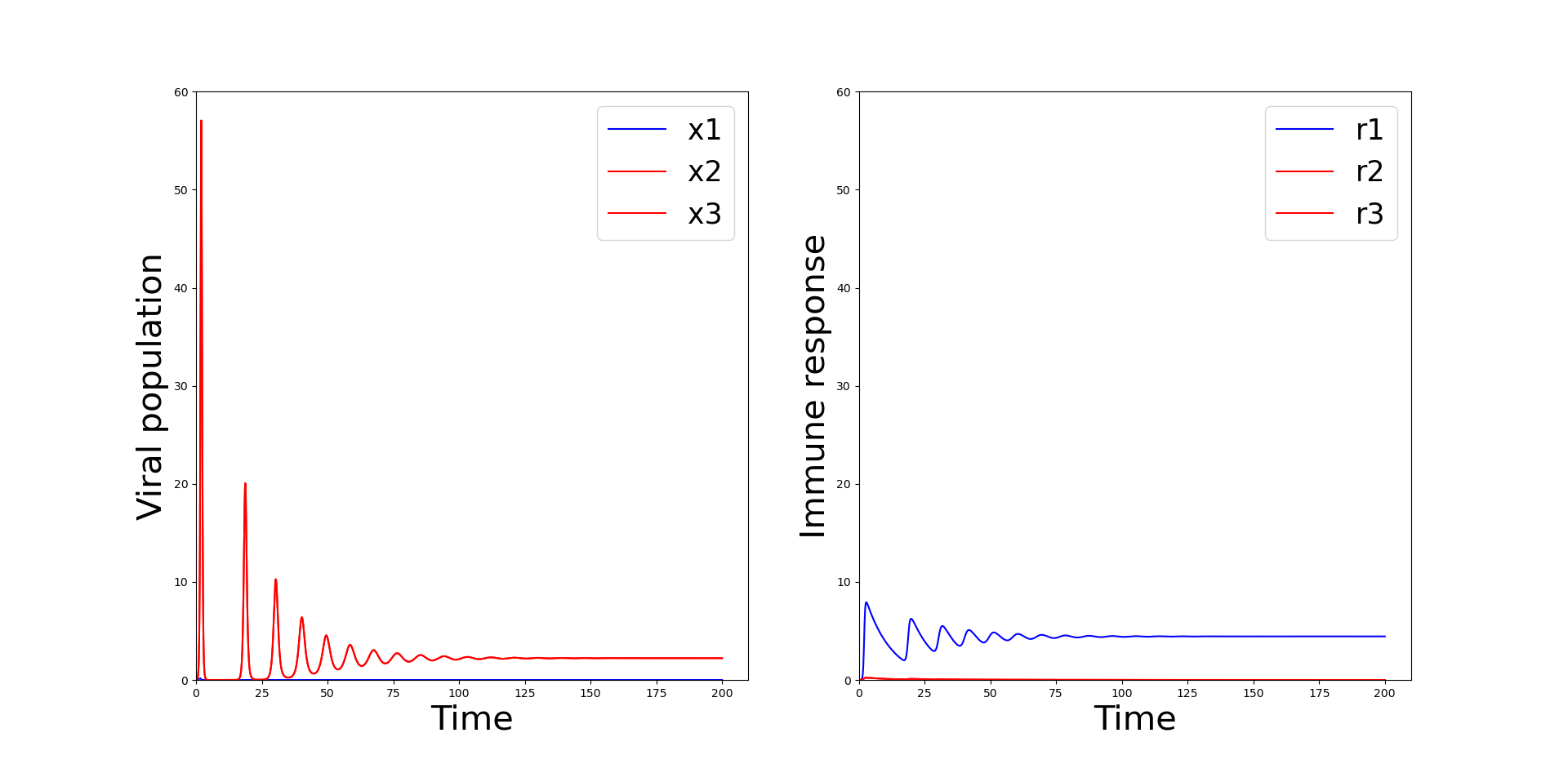}
    \caption{Numerical simulation of dynamics of the network shown in figure \ref{fig:simple_structure_one_altruist} with two persistent viruses. Virus 1 is altruistic (blue), and the others are persistent (red). The persistent viruses have the same population as a result of having the same replication rate and same types of connections. Figure produced using the parameters $f_1 = 2, f_2 = f_3 = 5, p = 2, c = 0.1, b = 0.1, \alpha = 3/4, K = 2, \beta = \alpha^K$.}
    \label{fig:simple_structure_dynamics}
\end{figure}

The results of computer simulations with the system (\ref{eqn:3network_ode}) demonstrate the state of stable local immunodeficiency, where the both persistent viruses get hidden from the host's immune system,
i.e. the immune response against them is virtually zero. At the same time, the immune response against the altruistic virus is essential while this virus virtually disappear, i.e. its concentration (size of population) becomes practically zero. Such evolution of this CR network is presented, for a specific choice of parameters, in figure \ref{fig:simple_structure_dynamics}.

\begin{figure}[ht]
     \centering
     \begin{subfigure}[b]{0.48\textwidth}
         \centering
         \includegraphics[width=\textwidth]{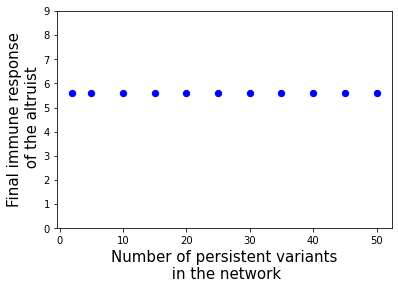}
         \caption{Final immune response of the altruistic virus}
         \label{fig:immune_response_multiple_persistent}
     \end{subfigure}
     \hfill
     \begin{subfigure}[b]{0.48\textwidth}
         \centering
         \includegraphics[width=\textwidth]{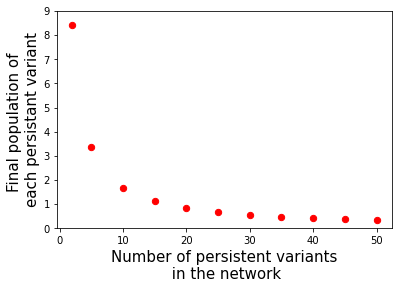}
         \caption{Final population of each persistent viruses}
         \label{fig:virus_pop_multiple_persistent}
     \end{subfigure}
        \caption{Results from simulating networks with a range of persistent viruses. The immune response of the altruist is unchanged when there are more persistent viruses connected to it. But the population of each persistent viruses drop when more variants are introduced.}
        \label{fig:final_immune_and_pop}
\end{figure}

As the goal is to find the maximum number of viruses that can be supported by an altruist, we tried adding more persistent viruses to the network shown in figure \ref{fig:simple_structure_one_altruist} using the same types of connections. Figure \ref{fig:final_immune_and_pop} shows how the population of persistent viruses and immune response of the altruistic virus change when more persistent variants are introduced to the network. The immune response of the altruist stays the same regardless of the number of variants connected to it in the CR network. But the population of each of the persistent variants decrease as the number of persistent variants introduced to the CR network increases. This indicates that practically there is an upper limit to the total number of viruses that can be supported by an altruistic virus. Indeed, each time as a new persistent virus is added equilibrium concentrations of persistent viruses decrease. However, practically there is a threshold, so that a virus cannot be detected (by immune system or a laboratory device) if its concentration goes below this threshold.   

The fixed point of the system \ref{eqn:3network_ode} that we were interested in is:
\begin{align*}
    x_1 &= 0, &x_2 &= \frac{b f_p}{2 \beta c p}, &x_3 &= \frac{b f_p}{2 \beta c p},\\
    r_1 &= \frac{f_p}{\beta p}, &r_2 &= 0, &r_3 &= 0.
\end{align*} 

Clearly this fixed point corresponds to the state of a local immunodeficiency where there are two positions (nodes) where LI takes place. This state of LI is stable. A proof of its stability can be found in appendix.

The major finding from the analysis of evolution of this system is that equilibrium concentrations of both persistent viruses are two times smaller than concentration of a single persistent virus connected to altruistic virus. Therefore a single altruistic virus seems can support only a fixed concentration of persistent viruses, which gets equally distributed if there are several persistent viruses in the state of stable LI. Our next goal will be to verify and demonstrate validity of this claim
numerically, and then to prove it analytically.

We will consider now the general case when a number of persistent
viruses in the network in figure \ref{fig:simple_structure_one_altruist} is arbitrary. We can look at a network with `$n$' viral variants (1 altruistic virus and $(n-1)$ persistent viruses connected to it) to generalize our conclusions. For the $n$-network, the dynamics of the network is:

\begin{flalign}\label{eqn:n_network_ode}
\begin{split}
    \dot{x_1} &= f_a x_1 - p x_1 r_1,\\
    \dot{x_2} &= f_p x_2 - p x_2 (\beta r_1 + r_2),\\
    &\cdots\\
    \dot{x_n} &= f_p x_n - p x_n (\beta r_1 + r_n),\\
    \dot{r_1} &= c\left(x_1 + \frac{\alpha x_2 r_1}{\alpha r_1 + r_2} + \cdots + \frac{\alpha x_n r_1}{\alpha r_1 + r_n}\right) - b r_1,\\
    \dot{r_2} &= c\left(\frac{x_2 r_2}{\alpha r_1 + r_2}\right) - b r_2,\\
    &\cdots\\
    \dot{r_n} &= c\left(\frac{x_n r_n}{\alpha r_1 + r_n}\right) - b r_n.
\end{split}
\end{flalign}
The fixed point that we were interested in is:
\begin{align*}
    x_1 &= 0 &x_2 &= \frac{b f_p}{n \beta c p} &\cdots& &x_n &= \frac{b f_p}{n \beta c p}\\
    r_1 &= \frac{f_p}{\beta p} &r_2 &= 0 &\cdots& &r_n &= 0
\end{align*}

These fixed points also show us that the final immune response of the altruist is independent on the number of persistent viruses attached to it, and hence it is a constant, as shown in figure \ref{fig:immune_response_multiple_persistent}.

We can use equations \ref{eqn:n_network_ode} to generalize that for a network with $n$ persistent viral variants and one altruist,\\
the final population of the persistent viral variant, $x_p = \frac{b f_p}{n \beta c p}$ and \\
the final immune response of the altruist $r_a = \frac{f_p}{\beta p}$.\\

As $n > 1$, $x_i^* = \cfrac{b f_p}{n \beta c p} < \cfrac{b f_p}{\beta c p}$ for every persistent variant $i$.

As $\beta < 1$, $r_1^* > \frac{f_p}{p}$, which is the immune response against the altruist.\\

We conclude this section by considering a weak (real) Local Immunodeficiency \cite{BS1}. It is impossible in numerical simulations to get to the state when populations of persistent viruses, which cannot be supported by altruists, become exactly zero. Likewise, it is impossible in numerics achieve the state when the immune response against persistent viruses becomes exactly zero. Therefore, in all numerical simulations (starting with \cite{pnas}) it was assumed that persistent or neutral viruses disappear if their concentration become less that their initial populations. However, in analytical rigorous studies we assume that these populations are identically zero. It would take an infinite time for population to get to this final stable state.

Another point to be clarified is that assuming that in the state of stable LI populations of altruistic viruses are identically zero, we may loose a connection with reality (biology in this case)
because if altruistic viruses are not present then they can not produce any effect. However, the (although highly nonlinear) equations (\ref{population}) governing the evolution of the system in question have the right hand sides which continuously depend on parameters.
Therefore, by for any small variation of parameters there exists a stable fixed point of our system. Particularly, there are positive but close to zero values of altruistic values concentrations in the state of the stable local immunodeficiency. Certainly, the corresponding values of strengths of immune responses against some persistent viruses also become not zeroes but small values. Therefore, there are many (practically non-distinguishable because they are very close to each other) real (weak) stable states of local immunodeficiency for our system. (One can find a more detailed analysis of this question in \cite{BS1,BS2}).     

Therefore, it follows from the results of this section that the maximum number $n$ of variants that can be supported by an altruistic virus is defined by the relation $\frac{b f_p}{n \beta c p} > x_i(0)$, where $x_i(0)$ is the initial value of persistent viruses $i$. i.e. in practical applications, the threshold should also depend on initial concentrations of persistent viruses. \\

\section{Maximal load on two altruistic viruses}

Our next goal is to understand whether and how altruistic viruses can cooperate in order to support more persistent viruses. In extensive numerical simulations conducted in \cite{pnas} always in equilibrium state altruistic viruses were connected to each other. Taking into account that it turned out that altruistic viruses comprise only about one percent of all types of viruses it is natural to question whether there is some kind of cooperation between altruistic viruses in order to support about ten times larger \cite{pnas} variety of persistent viruses. This important question has been never studied before.

In the present section we consider the simplest situation when there are only two altruistic viruses in a CR network. A goal is to determine the maximal load on two altruistic viruses and whether and how it changes depending on topology and the number of persistent viruses in a CR network. 

\begin{figure}[ht]
    \centering
    \includegraphics[width=0.5\textwidth]{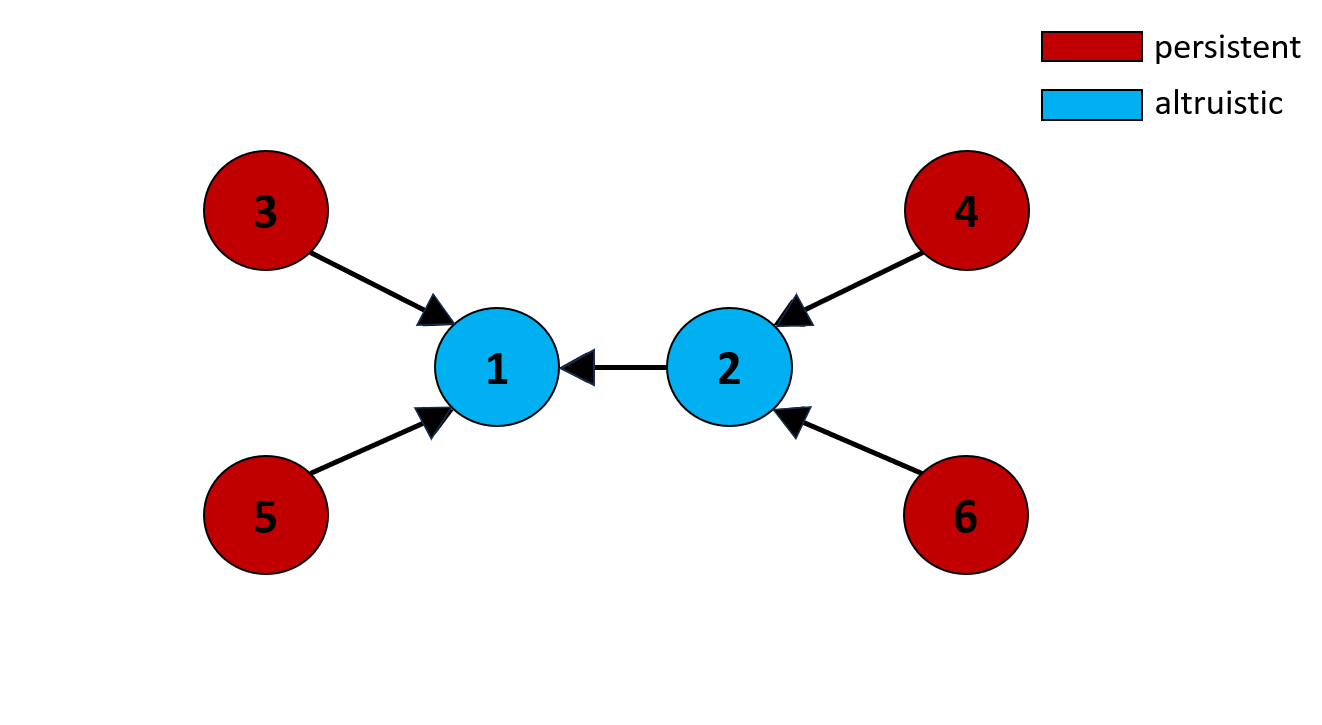}
    \caption{A basic network structure with two altruistic viruses and multiple persistent viruses}
    \label{fig:structure1}
\end{figure}

We start with a network which is, in a sense, similar to the one shown in figure \ref{fig:structure1}. In this network we have a directed connection (directed edge) between the altruistic viruses.

The dynamics of the network in figure \ref{fig:structure1} is described by the following system of ordinary differential equations
\begin{flalign}
\begin{split}
    \dot{x_1} &= f_a x_1 - p x_1 r_1,\\
    \dot{x_2} &= f_a x_2 - p x_2 (\beta r_1 + r_2),\\
    \dot{x_3} &= f_p x_3 - p x_3 (\beta r_1 + r_3),\\
    \dot{x_4} &= f_p x_4 - p x_4 (\beta r_2 + r_4),\\
    \dot{x_5} &= f_p x_5 - p x_5 (\beta r_1 + r_5)\\
    \dot{x_6} &= f_p x_6 - p x_6 (\beta r_2 + r_6),\\
    \dot{r_1} &= c(x_1 + \frac{\alpha x_2 r_1}{\alpha r_1 + r_2} + \frac{\alpha x_3 r_1}{\alpha r_1 + r_3}) + \frac{\alpha x_5 r_1}{\alpha r_1 + r_5}) - b r_1,\\
    \dot{r_2} &= c(\frac{x_2 r_2}{\alpha r_1 + r_2} + \frac{\alpha x_4 r_2}{\alpha r_2 + r_4} + \frac{\alpha x_6 r_2}{\alpha r_2 + r_6}) - b r_2,\\
    \dot{r_3} &= c(\frac{x_3 r_3}{\alpha r_1 + r_3}) - b r_3,\\
    \dot{r_4} &= c(\frac{x_4 r_4}{\alpha r_2 + r_4}) - b r_4,\\
    \dot{r_5} &= c(\frac{x_5 r_5}{\alpha r_1 + r_5}) - b r_5,\\
    \dot{r_6} &= c(\frac{x_6 r_6}{\alpha r_2 + r_6}) - b r_6.
\end{split}
\end{flalign}
The fixed point of this system is defined by the following relations

\begin{align*}
    x_1 &= 0 &x_2 &= 0 &x_3 &= x_4 = x_5 = x_6 = \frac{b f_p}{2 \beta c p}\\
    r_1 &= \frac{f_p}{\beta p} &r_2 &= \frac{f_p}{\beta p} &r_3 &= r_4 = r_5 = r_6 = 0
\end{align*}
This fixed point is stable under the conditions $c<1$ and $\alpha > c/2$.

One can see that it is actually a collection of two  fixed points which we got for a network with one altruist and two persistent viruses. Therefore the maximal load of a single altruistic virus does not change when connected to another altruistic virus. Essentially the result is the same as in case when two altruistic viruses are not connected.

\begin{figure}[h!]
    \centering
    \includegraphics[width=0.5\textwidth]{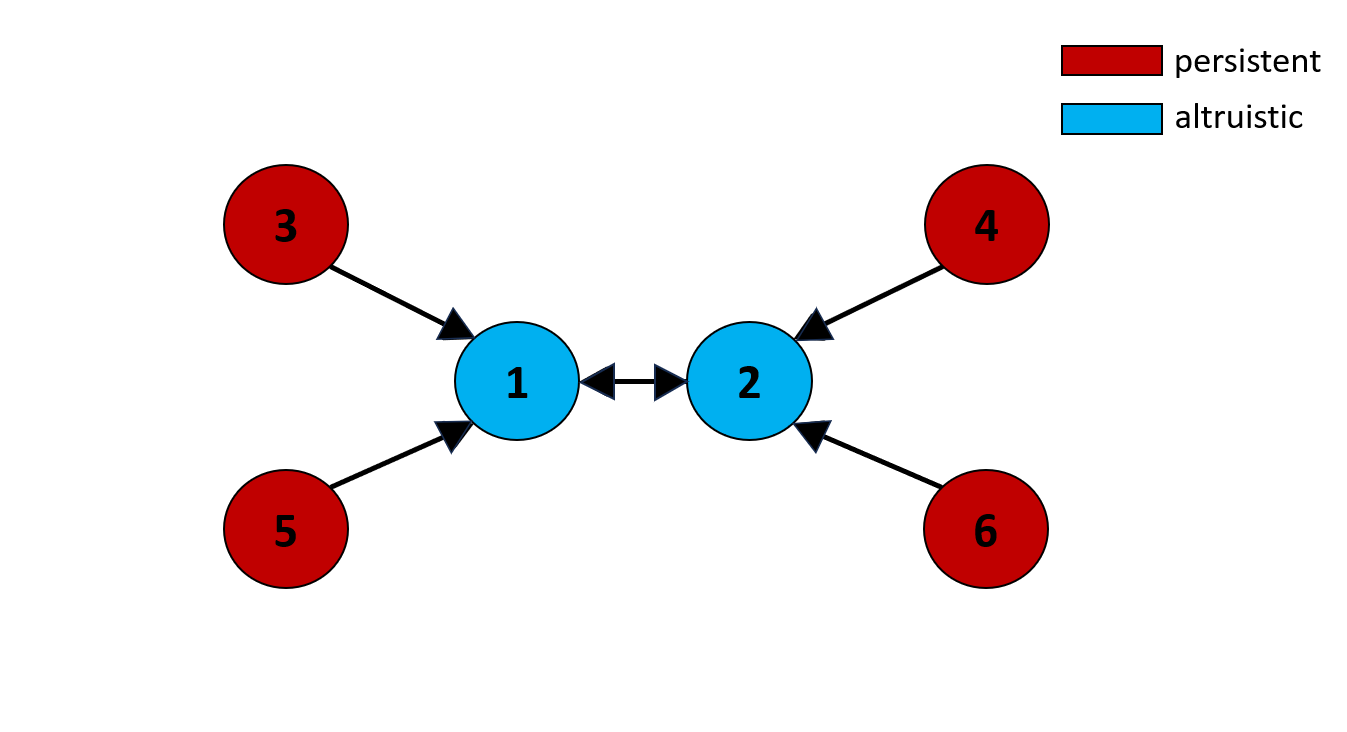}
    \caption{Two altruists supporting two persistent viruses each, with both altruists connected to each other}
    \label{fig:structure2}
\end{figure}

Next we analyze whether the symmetric connections between the two altruists (i.e. having an undirected edge between them) would result in any change (figure \ref{fig:structure2}).

The dynamics of the network shown in figure \ref{fig:structure2} is govern by the following equations
\begin{flalign}
\begin{split}
    \dot{x_1} &= f_a x_1 - p x_1(r_1 + \beta r_2),\\
    \dot{x_2} &= f_a x_2 - p x_2 (\beta r_1 + r_2),\\
    \dot{x_3} &= f_p x_3 - p x_3 (\beta r_1 + r_3),\\
    \dot{x_4} &= f_p x_4 - p x_4 (\beta r_2 + r_4)\\
    \dot{x_5} &= f_p x_5 - p x_5 (\beta r_1 + r_5)\\
    \dot{x_6} &= f_p x_6 - p x_6 (\beta r_2 + r_6),\\
    \dot{r_1} &= c(\frac{x_1 r_1}{r_1 + \alpha r_2} + \frac{\alpha x_2 r_1}{\alpha r_1 + r_2} + \frac{\alpha x_3 r_1}{\alpha r_1 + r_3}) + \frac{\alpha x_5 r_1}{\alpha r_1 + r_5}) - b r_1,\\
    \dot{r_2} &= c(\frac{\alpha x_1 r_2}{r_1 + \alpha r_2} + \frac{x_2 r_2}{\alpha r_1 + r_2} + \frac{\alpha x_4 r_2}{\alpha r_2 + r_4} + \frac{\alpha x_6 r_2}{\alpha r_2 + r_6}) - b r_2,\\
    \dot{r_3} &= c(\frac{x_3 r_3}{\alpha r_1 + r_3}) - b r_3,\\
    \dot{r_4} &= c(\frac{x_4 r_4}{\alpha r_2 + r_4}) - b r_4,\\
    \dot{r_5} &= c(\frac{x_5 r_5}{\alpha r_1 + r_5}) - b r_5,\\
    \dot{r_6} &= c(\frac{x_6 r_6}{\alpha r_2 + r_6}) - b r_6.
\end{split}
\end{flalign}
The fixed point of this system is 
\begin{align*}
    x_1 &= 0 &x_2 &= 0 &x_3 &= x_4 = x_5 = x_6 = \frac{b f_p}{2 \beta c p}\\
    r_1 &= \frac{f_p}{\beta p} &r_2 &= \frac{f_p}{\beta p} &r_3 &= r_4 = r_5 = r_6 = 0
\end{align*}

Again, we see that there is no difference with respect to the fixed points among the networks in figures \ref{fig:structure1} and \ref{fig:structure2}. 

Therefore any connection between the two altruists does not help each other in any way, when it comes to the capacity of persistent viruses that they can carry.

\begin{figure}[h!]
    \centering
    \includegraphics[width=0.5\textwidth]{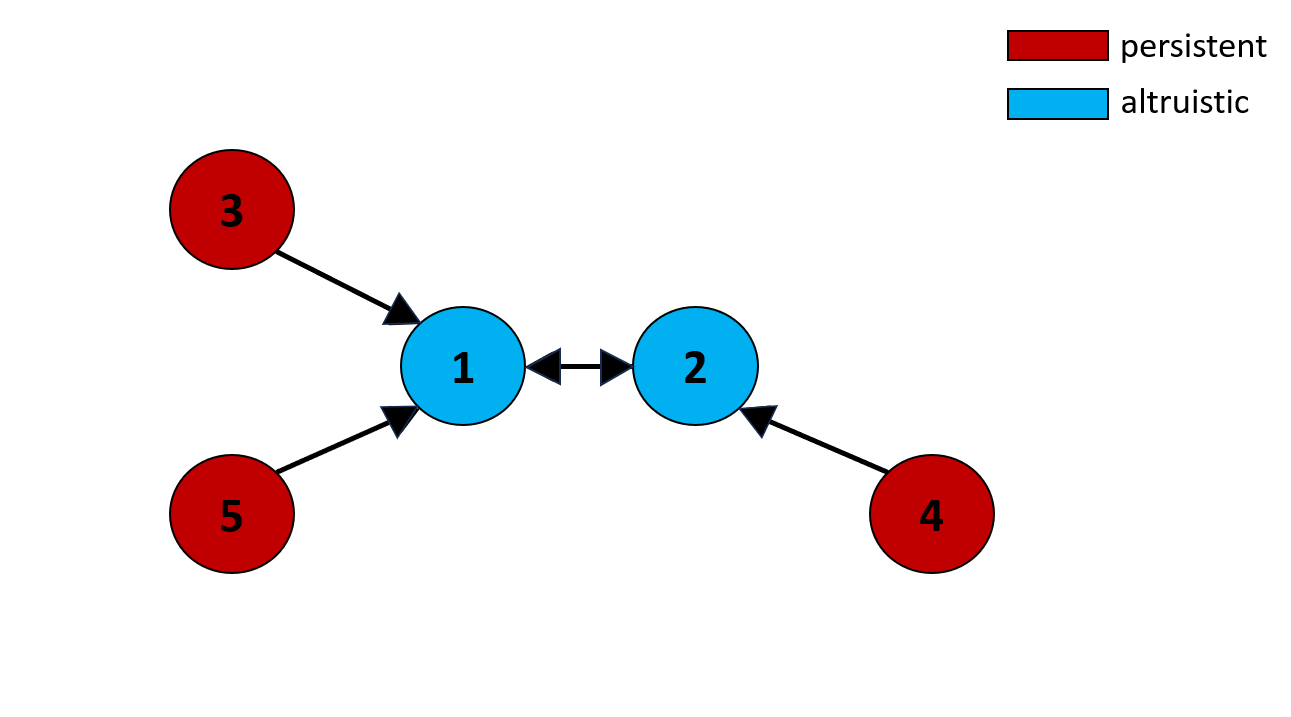}
    \caption{Two altruists supporting different number of persistent viruses}
    \label{fig:structure3}
\end{figure}

Next, to see if the maximal load of the two connected altruistic viruses are independent of each other, we considered the network shown in figure \ref{fig:structure3}.
The dynamics of the network in the figure \ref{fig:structure3} is governed by the following equations

\begin{flalign}
\begin{split}
    \dot{x_1} &= f_a x_1 - p x_1(r_1 + \beta r_2),\\
    \dot{x_2} &= f_a x_2 - p x_2 (\beta r_1 + r_2),\\
    \dot{x_3} &= f_p x_3 - p x_3 (\beta r_1 + r_3)\\
    \dot{x_4} &= f_p x_4 - p x_4 (\beta r_2 + r_4)\\
    \dot{x_5} &= f_p x_5 - p x_5 (\beta r_1 + r_5),\\
    \dot{r_1} &= c(\frac{x_1 r_1}{r_1 + \alpha r_2} + \frac{\alpha x_2 r_1}{\alpha r_1 + r_2} + \frac{\alpha x_3 r_1}{\alpha r_1 + r_3}) + \frac{\alpha x_5 r_1}{\alpha r_1 + r_5}) - b r_1,\\
    \dot{r_2} &= c(\frac{\alpha x_1 r_2}{r_1 + \alpha r_2} + \frac{x_2 r_2}{\alpha r_1 + r_2} + \frac{\alpha x_4 r_2}{\alpha r_2 + r_4}) - b r_2,\\
    \dot{r_3} &= c(\frac{x_3 r_3}{\alpha r_1 + r_3}) - b r_3,\\
    \dot{r_4} &= c(\frac{x_4 r_4}{\alpha r_2 + r_4}) - b r_4,\\
    \dot{r_5} &= c(\frac{x_5 r_5}{\alpha r_1 + r_5}) - b r_5.
\end{split}
\end{flalign}

The fixed point of this system is
\begin{align*}
    x_1 &= 0 &x_2 &= 0 &x_3 &=x_5 = \frac{b f_p}{2 \beta c p} &x_4 &=  \frac{b f_p}{\beta c p}\\
    r_1 &= \frac{f_p}{\beta p} &r_2 &= \frac{f_p}{\beta p} &r_3 &= r_5 = 0 &r_4 &= 0
\end{align*}

\noindent
We can see that persistent viruses 3 and 5, connected to altruist 1 have the equilibrium values $\frac{b f_p}{2 \beta c p}$, while the persistent virus 4, connected to altruist 2, has the equilibrium value which is twice as large $\frac{b f_p}{\beta c p}$. 

These results demonstrate that the fixed point depends only on the number of persistent viruses that are connected to that particular altruist. In other words, altruistic viruses, either connected or not, operate autonomously from each other. 

\begin{figure}[ht]
    \centering
    \includegraphics[width=0.5\textwidth]{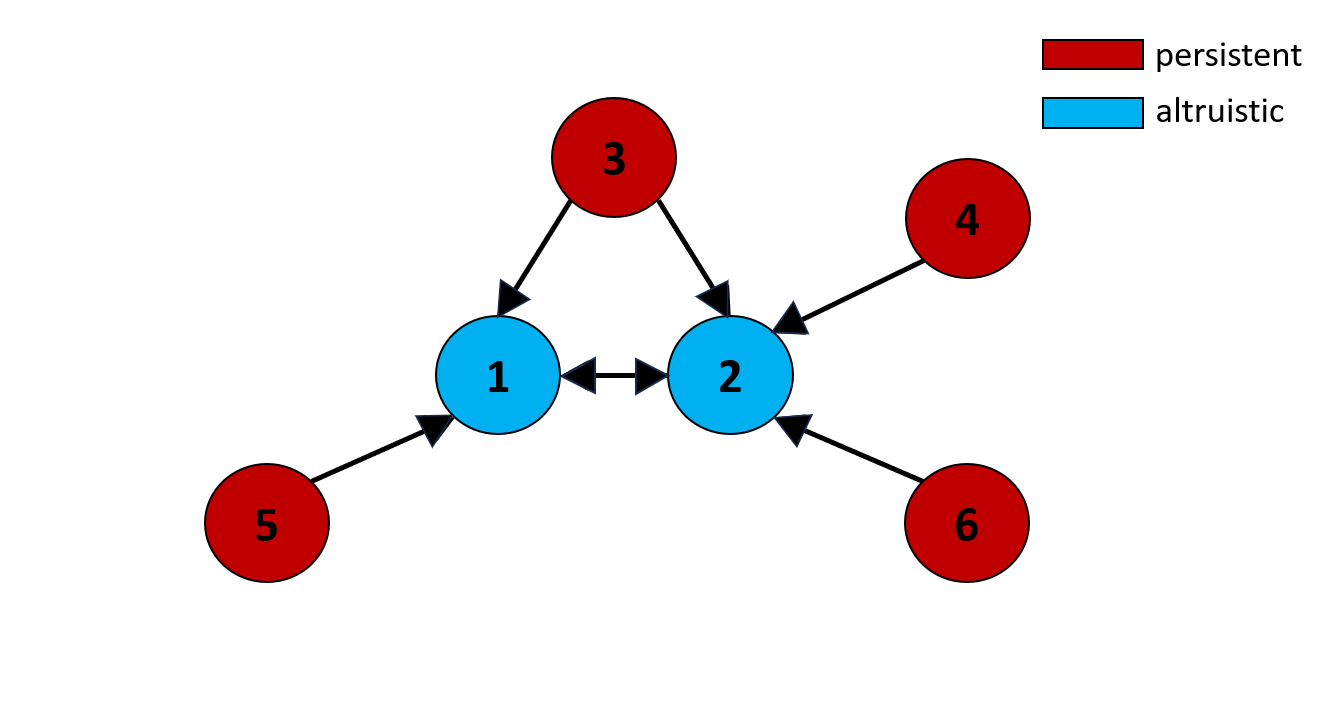}
    \caption{Two altruists supporting four persistent viruses, where one of the persistent viruses is connected to both the altruists}
    \label{fig:two_altruist_assymmetric}
\end{figure}

Now we are considering a case where in network \ref{fig:structure1}, one of the persistent viruses is supported by both altruists. The system of equations corresponding to this network (figure \ref{fig:two_altruist_assymmetric}) is the following:

\begin{flalign}
\begin{split}
    \dot{x_1} &= f_a x_1 - p x_1(r_1 + \beta r_2),\\
    \dot{x_2} &= f_a x_2 - p x_2 (\beta r_1 + r_2),\\
    \dot{x_3} &= f_3 x_3 - p x_3 (\beta r_1 + \beta r_2 + r_3),\\
    \dot{x_4} &= f_p x_4 - p x_4 (\beta r_2 + r_4)\\
    \dot{x_5} &= f_p x_5 - p x_5 (\beta r_1 + r_5)\\
    \dot{x_6} &= f_p x_6 - p x_6 (\beta r_2 + r_6),\\
    \dot{r_1} &= c\left(\frac{x_1 r_1}{r_1 + \alpha r_2} + \frac{\alpha x_2 r_1}{\alpha r_1 + r_2} + \frac{\alpha x_3 r_1}{\alpha r_1 + \alpha r2 + r_3} + \frac{\alpha x_5 r_1}{\alpha r_1 + r_5}\right) - b r_1,\\
    \dot{r_2} &= c\left(\frac{\alpha x_1 r_2}{r_1 + \alpha r_2} + \frac{x_2 r_2}{\alpha r_1 + r_2} + \frac{\alpha x_3 r_2}{\alpha r_1 + \alpha r2 + r_3} + \frac{\alpha x_4 r_2}{\alpha r_2 + r_4} + \frac{\alpha x_6 r_2}{\alpha r_2 + r_6}\right) - b r_2,\\
    \dot{r_3} &= c\left(\frac{x_3 r_3}{\alpha r_1 + \alpha r2 + r_3}\right) - b r_3,\\
    \dot{r_4} &= c\left(\frac{x_4 r_4}{\alpha r_2 + r_4}\right) - b r_4,\\
    \dot{r_5} &= c\left(\frac{x_5 r_5}{\alpha r_1 + r_5}\right) - b r_5,\\
    \dot{r_6} &= c\left(\frac{x_6 r_6}{\alpha r_2 + r_6}\right) - b r_6.
\end{split}
\end{flalign}
In this system, $f_a$ is the replication rate of the altruists, $f_3$ the replication rate of the shared persistent virus 3, and $f_p$ is the replication rate of the rest of the persistent viruses (viruses 4, 5, and 6).\\
To maintain the required dynamic with stability (where viruses 1, 2 are altruistic and the rest are persistent), we need $f_3 = 2 f_p$, i.e. the replication rate of the shared persistent to be twice that of the others.

The fixed point for this dynamic is described by the following expressions:\\

\begin{equation*}
    \begin{split}
        r_1 = r_2 = \frac{f_p}{\beta p},\\
        x_4 = x_6 = \frac{x_5}{2},\\
        \frac{x_3}{2} + x_5 = \frac{b f_p}{\beta c p},\\
        x_1 = x_2 = 0, r_3 = r_4 = r_5 = r_6 = 0.
    \end{split}
\end{equation*}

This fixed point is unchanged if the connection between the altruists are bidirectional or unidirectional, or if the altruists are completely disconnected.

\section{Discussion}
The findings of this study provide significant insights into the dynamics of viral populations within cross-immunoreactivity networks (CRNs), particularly focusing on the roles and interactions between altruistic and persistent viral variants. The results underscore several key points that contribute to our understanding of local immunodeficiency (LI) and its implications for viral evolution and immune system interactions.

One of the most notable outcomes of this research is the confirmation that altruistic viruses operate autonomously within CRNs. The connections among altruistic viruses do not alter their roles or the strengths of their interactions. Each altruistic virus functions independently, maintaining the same actions and interaction strengths regardless of the presence of other altruistic viruses. This independence suggests that the role of altruistic viruses in supporting persistent viruses is robust and does not rely on the network's complexity.

The presence of multiple altruistic viruses, however, contributes to maintaining the same population of persistent viruses. The current study extends this understanding by demonstrating that a CRN can achieve stable LI even with only persistent variants, challenging the previous assumption that altruistic variants are indispensable in the cooperative dynamics of infections like hepatitis-C.

The study addresses a critical question regarding the maximum number of persistent viruses that can be supported by an altruistic virus. In the computer simulations performed in \cite{pnas}, it was found that the number of persistent viruses connected to a single altruistic virus can significantly exceed the number of altruistic viruses by about ten times on average. This ratio underscores the efficiency of altruistic viruses in supporting persistent variants, despite the competitive nature of the persistent viruses. The simulations showed that as more persistent viruses are added to the network, the population of each persistent virus decreases, suggesting an upper limit to the number of persistent viruses an altruistic virus can support. This limit is crucial for understanding the balance and stability within viral populations and has implications for the design of antiviral strategies and vaccines. The study concludes that the maximum number $n$ of variants that can be supported by a single altruistic variant is defined by the relation $\frac{b f_p}{n\beta cp} > x_i(0)$, where $x_i(0)$ is the initial population of persistent variant $i$, and other parameters are as described in section \ref{sec: model}.

The study's insights into antigenic cooperation and the quasi-social behavior of viral variants have profound implications for vaccine design. The complexity of interactions within CRNs and the role of altruistic viruses in maintaining LI suggest that vaccines targeting these networks need to account for the dynamic and competitive nature of viral populations. Effective vaccines may need to disrupt the supportive roles of altruistic viruses or enhance the immune system's ability to target persistent variants directly.

The study does have some limitations. While the model analyzed is fairly detailed, it is not exhaustive and does not encompass several immunological aspects, such as T cell immunity and antibody competition for neutralization (as opposed to the competition for activation, which the model considers). Future research should explore more complex CRNs with varying degrees of connectivity and diversity among viral variants. Understanding how different network structures affect the stability of LI and the evolution of viral populations can provide deeper insights into the mechanisms of viral evolution. Additionally, experimental validation of these theoretical models would be invaluable in confirming the predicted behaviors and interactions within CRNs.

In conclusion, this study advances our understanding of the intricate dynamics within viral CRNs, highlighting the crucial roles of altruistic and persistent viruses. The findings provide a foundation for developing more effective antiviral strategies and contribute to the broader knowledge of viral evolution and immune system interactions.

\appendix
\section{Appendix}

\subsection{Stability of the fixed point of 3-cycle network}\label{3-cycle}
The Jacobian of the system of  (\ref{eqn:3-cycle}) at the fixed point shown in figure \ref{fig:3-cycle} is $J = (X Y)$ where:
$$
X = \scalemath{0.7}{\left(\begin{array}{cccc} f_{1}-p\,\left(r_{1}+\beta \,r_{2}\right) & 0 & 0 & -p\,x_{1}\\ 0 & f_{2}-p\,\left(r_{2}+\beta \,r_{3}\right) & 0 & 0\\ 0 & 0 & f_{3}-p\,\left(r_{3}+\beta \,r_{1}\right) & -\beta \,p\,x_{3}\\ \frac{c\,r_{1}}{r_{1}+\alpha \,r_{2}} & 0 & \frac{\alpha \,c\,r_{1}}{r_{3}+\alpha \,r_{1}} & c\,\left(\frac{x_{1}}{r_{1}+\alpha \,r_{2}}+\frac{\alpha \,x_{3}}{r_{3}+\alpha \,r_{1}}-\frac{r_{1}\,x_{1}}{{\left(r_{1}+\alpha \,r_{2}\right)}^2}-\frac{\alpha ^2\,r_{1}\,x_{3}}{{\left(r_{3}+\alpha \,r_{1}\right)}^2}\right)-b\\ \frac{\alpha \,c\,r_{2}}{r_{1}+\alpha \,r_{2}} & \frac{c\,r_{2}}{r_{2}+\alpha \,r_{3}} & 0 & -\frac{\alpha \,c\,r_{2}\,x_{1}}{{\left(r_{1}+\alpha \,r_{2}\right)}^2}\\ 0 & \frac{\alpha \,c\,r_{3}}{r_{2}+\alpha \,r_{3}} & \frac{c\,r_{3}}{r_{3}+\alpha \,r_{1}} & -\frac{\alpha \,c\,r_{3}\,x_{3}}{{\left(r_{3}+\alpha \,r_{1}\right)}^2} \end{array}\right)
}$$
and
$$
Y = \scalemath{0.7}{
\left(\begin{array}{cc} -\beta \,p\,x_{1} & 0\\ -p\,x_{2} & -\beta \,p\,x_{2}\\ 0 & -p\,x_{3}\\ -\frac{\alpha \,c\,r_{1}\,x_{1}}{{\left(r_{1}+\alpha \,r_{2}\right)}^2} & -\frac{\alpha \,c\,r_{1}\,x_{3}}{{\left(r_{3}+\alpha \,r_{1}\right)}^2}\\ c\,\left(\frac{x_{2}}{r_{2}+\alpha \,r_{3}}+\frac{\alpha \,x_{1}}{r_{1}+\alpha \,r_{2}}-\frac{r_{2}\,x_{2}}{{\left(r_{2}+\alpha \,r_{3}\right)}^2}-\frac{\alpha ^2\,r_{2}\,x_{1}}{{\left(r_{1}+\alpha \,r_{2}\right)}^2}\right)-b & -\frac{\alpha \,c\,r_{2}\,x_{2}}{{\left(r_{2}+\alpha \,r_{3}\right)}^2}\\ -\frac{\alpha \,c\,r_{3}\,x_{2}}{{\left(r_{2}+\alpha \,r_{3}\right)}^2} & c\,\left(\frac{x_{3}}{r_{3}+\alpha \,r_{1}}+\frac{\alpha \,x_{2}}{r_{2}+\alpha \,r_{3}}-\frac{r_{3}\,x_{3}}{{\left(r_{3}+\alpha \,r_{1}\right)}^2}-\frac{\alpha ^2\,r_{3}\,x_{2}}{{\left(r_{2}+\alpha \,r_{3}\right)}^2}\right)-b \end{array}\right)
}$$
With the parameters $f = 2, c = 1, p = 2, \alpha = 2/3, \beta = 4/9, b = 3$ that satisfy these conditions, the eigenvalues of the system at this fixed point are:
\begin{flalign*}
  &{\lambda_1 = -0.60374 + 1.3052i} &\quad   &{\lambda_2 = -0.60374 - 1.3052i}\\
  &{\lambda_3 = -0.23625 + 1.3052i} &\quad   &{\lambda_4 = -0.23625 - 1.3052i}\\
  &{\lambda_5 = -1.5 + 1.9365i} &\quad   &{\lambda_6 = -1.5 - 1.9365i}
\end{flalign*}
Hence we can generate a system with the CRN in figure \mbox{\ref{fig:3-cycle}} which has a stable and robust steady state of LI.\\

\subsection{Stability of the fixed point of basic network with two persistent and one altruistic virus}\label{3network_basic}
The system of equations \ref{eqn:3network_ode} showed the case where a basic network has two persistent and one altruistic virus. At the fixed points mentioned, the Jacobian of the system would be as follows:

$$
J = \left(\begin{array}{cccccc} f_{1}-\frac{\mathrm{fp}}{\beta } & 0 & 0 & 0 & 0 & 0\\ 0 & 0 & 0 & -\frac{b\,\mathrm{fp}}{2\,c} & -\frac{b\,\mathrm{fp}}{2\,\beta \,c} & 0\\ 0 & 0 & 0 & -\frac{b\,\mathrm{fp}}{2\,c} & 0 & -\frac{b\,\mathrm{fp}}{2\,\beta \,c}\\ c & c & c & -b & -\frac{b}{2\,\alpha } & -\frac{b}{2\,\alpha }\\ 0 & 0 & 0 & 0 & -\frac{b\,\left(2\,\alpha -1\right)}{2\,\alpha } & 0\\ 0 & 0 & 0 & 0 & 0 & -\frac{b\,\left(2\,\alpha -1\right)}{2\,\alpha } \end{array}\right).
$$
The eigenvalues of this system can be calculated to be:

$$
\lambda = \left(\begin{array}{c} 0\\ -\frac{b}{2}-2\,\sqrt{\frac{b\,\left(b-4\,f_p\right)}{16}}\\ 
-\frac{b}{2}+2\,\sqrt{\frac{b\,\left(b-4\,f_p\right)}{16}}\\
-\frac{b(2\alpha-1)}{2\,\alpha }\\
-\frac{b(2\alpha-1)}{2\,\alpha }\\
-\frac{f_p-\beta \,f_{1}}{\beta } \end{array}\right)
$$
We will present now an exact example with the stable state of local immunodeficiency. Let the system’s parameters have the following values $f_1 = 0.25, f_p = 0.3, c = 1, p = 2, \alpha = 2/3, \beta = 1/9, b = 3$. For these values, the eigenvalues of the system are non-positive and are hence the fixed point is stable.


\bibliographystyle{elsarticle-num} 
\bibliography{ref}

\end{document}